\documentclass[aps]{revtex4}
\usepackage{amsmath,graphicx}
\usepackage{amssymb}
\usepackage{color}
\newcommand{\be}{\begin{equation}}
\newcommand{\ee}{\end{equation}}
\newcommand{\ba}{\begin{eqnarray}}
\newcommand{\ea}{\end{eqnarray}}
\newcommand{\br}{\mathbf r}
\newcommand{\bq}{\mathbf q}
\newcommand{\e}{\varepsilon}
\DeclareMathOperator{\sgn}{sgn}
\DeclareMathOperator{\tr}{tr}
\DeclareMathOperator{\re}{\rm Re}
\DeclareMathOperator{\im}{\rm Im}

\begin{document}

\title{Dynamical polarization  of monolayer graphene  in a magnetic field}

\author{P. K. Pyatkovskiy}
\affiliation{Department of Applied Mathematics, University of Western Ontario,
London, Ontario N6A 5B7, Canada}

\author{V. P. Gusynin}
\affiliation{Bogolyubov Institute for Theoretical Physics, 03680, Kiev, Ukraine}

\begin{abstract}
The one-loop dynamical polarization function of graphene in an external magnetic
field is calculated as a function of wavevector and frequency
at finite chemical potential, temperature, band gap, and width of Landau levels.
The exact analytic result is given in terms of digamma functions and generalized
Laguerre polynomials, and has the form of double sum over Landau levels. Various
limits (static, clean, etc) are discussed. The Thomas-Fermi
inverse length $q_F$ of screening  of the Coulomb potential  is
found to be an oscillating function of a magnetic field and a
chemical potential. At zero temperature and scattering rate, it
vanishes when the Fermi level lies between the Landau levels.
\end{abstract}

\maketitle

\section{Introduction}

The fabrication of graphene~\cite{Novoselov2004S} initiated extensive
theoretical and experimental studies of its remarkable electronic properties aimed
at promised applications of this material in next-generation electronic devices.
The non-interacting charge carriers in single layer graphene are described by the analogue
of the Dirac equation for the massless fermions with the relativistic-like linear
spectrum~\cite{Semenoff1984} and a vanishing density of states at zero doping.
In the presence of the external magnetic field the spectrum of
these Dirac quasiparticles has the form of relativistic Landau levels, in contrast to the
equidistantly spaced levels in a usual two-dimensional electron gas. These peculiar
features of the non-interacting charge carriers in graphene result in several interesting
physical phenomena such as the unconventional quantum Hall
effect~\cite{AndoPRB2002,GusyninPRL2005,PeresPRB2006,QHE-experiment}, the universal
optical conductivity~\cite{universal-optical-theory,universal-optical-exp}
and magneto-spectroscopy~\cite{PeresPRB2006,magneto-spectroscopy-theory,magneto-spectroscopy-exp}.

Although these and other electronic and transport phenomena in graphene are well
described in terms of free Dirac quasiparticles, the effects of interactions, in particular,
the Coulomb interaction, are not settled yet. The vanishing density of states
at the Dirac point ensures that the Coulomb interaction between the electrons remains
unscreened due to vanishing of the static polarization for $q\to0$ \cite{Gonzalez}.
The large value of the unscreened coupling constant $g=e^{2}/\hbar v_{F}$, where $e$
is the electron charge, $v_{F}\approx10^6$m/s is the Fermi velocity, could lead to instability in
pristine graphene and formation of excitonic condensate and a quasiparticle gap,
followed by quantum phase transition to an insulating phase above some critical $g_{c}$.
This possibility is studied in a series of theoretical works
\cite{Khveshchenko2001PRL,Gorbar2002PRB} (see, also recent papers \cite{instability})
but experimental evidence for such an insulating phase is still absent \cite{footnote}.

The screening of Coulomb potential due to the many-body interactions is determined
by the polarization function which is also an important physical quantity for
the spectrum of collective excitations (plasmons). This function in monolayer graphene
without a magnetic field has been studied in one-loop approximation
in~Refs.\cite{Gorbar2002PRB,Wunsch2006NJP,Pyatkovskiy2009JPCM}.
In the presence of an external magnetic field, it was calculated in~\cite{Shizuya2007PRB}
at zero temperature and impurity rate  with the result given by the double sum over
the Landau levels. The similar expression was also obtained later in~\cite{Roldan},
where it was employed to study the spectrum of collective excitations in a magnetic field.
However, to the best of our knowledge, the most general expression for the dynamical
polarization in the presence of finite temperature, chemical potential, impurity rate,
quasiparticle gap and a magnetic field was not given in the literature.

The present paper deals with this more general case.
The paper is organized as follows.
In Sec.~\ref{secII} we describe the model used and present our main
result for the polarization function.
We consider the clean graphene limit of this function in Sec.~\ref{secIII}.
In Sec.~\ref{SecIV} we focus on the static screening properties of graphene.
Then, in Sec.~\ref{SecV} we discuss some other limits of the polarization function,
and in Sec.~\ref{SecVI} we give the brief summary of our results.
Finally, we provide the details of the calculations in the appendices A and B.
In the appendix A we derive the expression for the dynamical polarization as a double sum
over the Landau levels while in the appendix B we employ the Schwinger proper time method
to get a double integral representation for the polarization.

\section{Model and general expression for polarization function}
\label{secII}

The Lagrangian describing the non-interacting Dirac quasiparticles
confined to the graphene plane, in an external magnetic field, reads
(we use the units $\hbar=c=1$)
\be
\mathcal L=\sum_{\sigma=1}^{N_f}\bar\Psi_\sigma\Bigl[i\gamma^0(\partial_t-i\mu)
+iv_F\boldsymbol\gamma(\boldsymbol\nabla+ie\mathbf A^{\rm ext})-\Delta\Bigr]\Psi_\sigma\,,
\ee
where $\Psi_\sigma^T=(\psi^\sigma_{KA},\psi^\sigma_{KB},\psi^\sigma_{K'B},\psi^\sigma_{K'A})$
is the four-component wave function describing the Bloch states on the $A$ and $B$ sublattices
and in the vicinity of $\mathrm K$ and $\mathrm K'$ points in the momentum space.
$\bar\Psi_\sigma=\Psi_\sigma^\dag\gamma^0$ is the Dirac conjugated spinor,
$\sigma$ is the spin variable, and gamma-matrices $\gamma^\nu=\sigma_3
\otimes(\sigma_3,i\sigma_2,-i\sigma_1)$ form the reducible $4\times4$ representation
in $2+1$ dimensions.

We will neglect the Zeeman splitting which in graphene is very small ($\sim1.34B$[T] ${K}$)
compared to the distance between the zeroth and the first Landau levels ($\sim424\sqrt{B[{\rm T}]}$ K).
Therefore, the electron spin results in only the degeneracy factor (number of flavors) $N_f=2$.
We have also included the gap term $\Delta$ which can be induced
in graphene by placing it on a top of an appropriate substrate \cite{substrateind-gap}
that breaks the sublattice symmetry, or can be generated dynamically in magnetic field
(the phenomenon of magnetic catalysis)~\cite{Khveshchenko2001PRL,Gorbar2002PRB}.
The external magnetic field $\mathbf B=\nabla\times\mathbf A^{\rm ext}$ is applied normally
to the graphene plane and the vector potential is taken in the symmetric gauge
$\mathbf A^{\rm ext}=(-By/2,Bx/2)$. The chemical potential $\mu$ can be varied
by applying the gate voltage.

The Green's function of Dirac quasiparticles described by this Lagrangian in
an external magnetic field reads
\be
G(t-t',\br;\br')=\exp\bigl(-ie\br\mathbf A^{\rm ext}(\br')\bigr)
S(t-t',\br-\br')\,,
\ee
where $S(t-t',\br-\br')$ is the translation invariant part of the propagator.
Using the expression for $S(i\omega_s,\bq)$ from~\cite{Chodos1990PRD,Gorbar2002PRB}
we obtain for the propagator in the configuration space (in the Matsubara representation)
\be
\label{prop}
S(i\omega_m,\br)=\frac i{2\pi l^2}\exp\Bigl(-\frac{\br^2}{4l^2}\Bigr)\sum_{n=0}^{\infty}
\frac{[\gamma^0(i\omega_m+\mu+i\Gamma_n\sgn\omega_m)+\Delta]f_1^n(\br)+f_2^n(\br)}
{(i\omega_m+\mu+i\Gamma_n\sgn\omega_m)^2-M_n^2}\,,\qquad
\omega_m=(2m+1)\pi T\,,
\ee
where $T$ is the temperature (we use $k_{\rm B}=1$), $M_n=\sqrt{2nv_F^2/l^2+\Delta^2}$,
and $E_{n}=\pm M_{n}$ are the energies of the relativistic Landau levels, $l=1/\sqrt{|eB|}$
is the magnetic length. The functions $f_{1,2}^n(\br)$ are defined as
\ba
f_1^n(\br)&=&P_-L_n\Bigl(\frac{\br^2}{2l^2}\Bigr)
+P_+L_{n-1}\Bigl(\frac{\br^2}{2l^2}\Bigr)\,,
\qquad P_\pm=\frac{1}{2}\left(1\pm i\gamma^1\gamma^2\sgn(eB)\right)\\
f_2^n(\br)&=&-\frac{iv_F}{l^2}(\boldsymbol\gamma\cdot\br)L_{n-1}^1\Bigl(\frac{\br^2}{2l^2}\Bigr),
\ea
where $L_n^\alpha(z)$ are the generalized Laguerre polynomials
(by definition, $L_n(z)\equiv L_n^0(z)$ and $L_{-1}^\alpha(z)\equiv0$).

The finite parameter $\Gamma_n$ has the meaning of the width of Landau levels or, equivalently,
the scattering rate of Dirac quasiparticles.
It is expressed through the retarded fermion self energy and in general depends
on the energy, temperature, magnetic field, and the Landau level index.
In our calculations, we are assume that the width is independent
of the energy (frequency).

The dynamical polarization determines many physically interesting properties, such
as the effective electron-electron interaction, the Friedel oscillations and
the spectrum of collective modes.
The retarded one-loop dynamical polarization function is given by the expression
\be
\label{pol_init}
\Pi(i\Omega_s,\bq)=e^2TN_f\int d^2r\,e^{-i\bq\br}
\sum_{m=-\infty}^\infty\tr\bigl[\gamma^0S(i\omega_m,\br)\gamma^0S(i\omega_m-i\Omega_s,-\br)\bigr]\,,
\qquad
\Omega_s=2\pi sT\,,
\ee
analytically continued from Matsubara frequencies to real $\Omega$ axis.
Note that our definition of the polarization function differs in the factor of $-e^2$
from that used in Refs.\cite{Shizuya2007PRB,Roldan}.
Details of the calculation of this function are given in the appendix,
here we reproduce only the final expression and then analyze various limiting cases.
Thus our main result reads
\be
\label{pol_main}
\begin{split}
\Pi(\Omega,\bq)=&\frac{e^2N_f}{4\pi l^2}
\sum_{n,n'=0}^{\infty}\sum_{\lambda,\lambda'=\pm}Q_{nn'}^{\lambda\lambda'}(y,\Delta)
\biggl[\frac{Z_{nn'}^{\lambda\lambda'}(\Omega,\Gamma,\mu,T)}
{\lambda M_n-\lambda'M_{n'}-\Omega-i(\Gamma_n-\Gamma_{n'})}\\
&+\frac{Z_{n'n}^{-\lambda',-\lambda}(\Omega,\Gamma,-\mu,T)}
{\lambda M_n-\lambda'M_{n'}-\Omega-i(\Gamma_{n'}-\Gamma_n)}
-\frac{Z_{nn}^{\lambda\lambda}(\Omega,\Gamma,\mu,T)
+Z_{n'n'}^{-\lambda',-\lambda'}(\Omega,\Gamma,-\mu,T)}
{\lambda M_n-\lambda'M_{n'}-\Omega-i(\Gamma_n+\Gamma_{n'})}\biggr]\,,
\end{split}
\ee
where we have introduced the following notations:
\be
\label{Z}
Z_{nn'}^{\lambda\lambda'}(\Omega,\Gamma,\mu,T)=\frac1{2\pi i}\biggl[
\psi\biggl(\frac12+\frac{\mu-\lambda M_n+\Omega+i\Gamma_n}{2i\pi T}\biggr)
-\psi\biggl(\frac12+\frac{\mu-\lambda'M_{n'}+i\Gamma_{n'}}{2i\pi T}\biggr)\biggr]\,,
\ee
and
\be
\label{Q}
\begin{split}
Q_{nn'}^{\lambda\lambda'}(y,\Delta)={}&e^{-y}y^{|n-n'|}\Biggl\{
\biggl(1+\frac{\lambda\lambda'\Delta^2}{M_nM_{n'}}\biggr)
\biggl(\frac{n_<!}{n_>!}\Bigl[L_{n_<}^{|n-n'|}(y)\Bigr]^2
+(1-\delta_{0n_<})\frac{(n_<-1)!}{(n_>-1)!}\Bigl[L_{n_<-1}^{|n-n'|}(y)\Bigr]^2\biggr)\\
&+\frac{4\lambda\lambda'v_F^2}{l^2M_nM_{n'}}\frac{n_<!}{(n_>-1)!}
L_{n_<-1}^{|n-n'|}(y)L_{n_<}^{|n-n'|}(y)\Biggr\}\,,
\end{split}
\ee
where $y=l^2\bq^2/2$, $n_>=\max(n,n')$, $n_<=\min(n,n')$,  and $\psi(z)$ is the digamma function.
Note the symmetry properties of the function $Q_{nn'}^{\lambda\lambda'}(y,\Delta)$
with respect to the exchange of indices $\lambda,\lambda'$ and $n,n'$ and
$Q_{nn'}^{\lambda\lambda'}(y,\Delta)=Q_{nn'}^{-\lambda,-\lambda'}(y,\Delta)$.

For the gapless graphene (with $\Delta=0$) the function~(\ref{Q}) reduces to
\be
\label{Q_gapless}
Q_{nn'}^{\lambda\lambda'}(y,0)={}e^{-y}y^{|n-n'|}\Biggl(
\sqrt{\frac{(1+\lambda\lambda'\delta_{0n_>})n_<!}{n_>!}}L_{n_<}^{|n-n'|}(y)
+\lambda\lambda'(1-\delta_{0n_<})
\sqrt{\frac{(n_<-1)!}{(n_>-1)!}}L_{n_<-1}^{|n-n'|}(y)\Biggr)^2\,.
\ee
Taking the limit of zero temperature, the expression~(\ref{Z}) simplifies to
\be
Z_{nn'}^{\lambda\lambda'}(\Omega,\Gamma,\mu,0)=\frac1{2\pi i}
\ln\biggl(\frac{\mu-\lambda M_n+\Omega+i\Gamma_n}{\mu-\lambda'M_{n'}+i\Gamma_{n'}}\biggr)\,.
\ee
The polarization function~(\ref{pol_main}) is an analytic function of $\Omega$
without singularities in the whole upper complex half-plane.
It depends only on the absolute value of chemical potential
(that can be verified by the replacement $\lambda\leftrightarrow-\lambda', n\leftrightarrow n'$)
and obeys the relation $\Pi(-\Omega,\bq)=[\Pi(\Omega,\bq)]^*$
(can be verified by the replacement $\lambda\leftrightarrow \lambda', n\leftrightarrow n'$
and taking into account
$[Z_{nn'}^{\lambda\lambda'}(\Omega,\Gamma,\mu,t)]^*=-Z_{nn'}^{-\lambda,-\lambda'}(-\Omega,\Gamma,-\mu,t)$).
At finite scattering rate, the polarization function~(\ref{pol_main})
receives the contributions both from the inter-
(with $\lambda n\ne\lambda'n'$) and the intra-Landau level ($\lambda n=\lambda'n'$) transitions.
Note that $Q_{00}^{\lambda,-\lambda}(y,\Delta)=0$ which reflects the fact that the levels
with energies $\pm\Delta$ belong to the different valleys,
and the intervalley transitions are not incorporated in our model.

\section{Clean graphene}
\label{secIII}

In the absence of scattering of Dirac quasiparticles ($\Gamma_n=0$) the general expression~(\ref{pol_main})
for the polarization function reduces by means of Eq.(\ref{splusr}) to the following form:
\be
\label{pol_clean}
\Pi(\Omega,\bq)=-\frac{e^2N_f}{4\pi l^2}\sum_{n,n'=0}^{\infty}\sum_{\lambda,\lambda'=\pm}
Q_{nn'}^{\lambda\lambda'}(y,\Delta)\frac{n_F(\lambda M_n)-n_F(\lambda'M_{n'})}
{\lambda M_n-\lambda'M_{n'}+\Omega+i0}\,,
\ee
where $n_F(x)=[e^{(x-\mu)/T}+1]^{-1}$.
One can easily see from the above expression that only the terms with $\lambda n\ne\lambda'n'$
(corresponding to the inter-Landau level transitions) survive in the clean limit. However, this is
not the case when the limit $\Gamma\to0$ is taken after setting $\Omega=0$
(see Eq.(\ref{pol_static_clean}) below).
When both scattering rate and temperature are zero, it simplifies further to
(the order of taking limits $\Gamma_n\to0$ and $T\to0$ is not important)
\be
\label{pol_Gamma0_T0}
\Pi(\Omega,\bq)=\frac{e^2N_f}{4\pi l^2}\sum_{n,n'=0}^{\infty}\sum_{\zeta=\pm}
\frac{Q_{nn'}^{-+}(y,\Delta)}{M_n+M_{n'}+\zeta(\Omega+i0)}
+\frac{e^2N_f}{4\pi l^2}
\theta(\mu^2-\Delta^2)\sum_{n=0}^{\infty}\sum_{n'=0}^{N_F}\sum_{\lambda,\zeta=\pm}
\frac{Q_{nn'}^{\lambda+}(y,\Delta)}
{\lambda M_n-M_{n'}+\zeta(\Omega+i0)}\,,
\ee
where we used the symmetry of the function $Q_{nn'}^{\lambda\lambda'}(y,\Delta)$
with respect to upper indices,
\be
N_F=\left[\frac{(\mu^2-\Delta^2)l^2}{2v_F^2}\right]
\ee
is the number of the highest filled Landau level (square brackets here denote
the integer part of expression). For $\mu<0$ it is a positive number meaning
the highest empty Landau level in the valence band. The first term
in Eq.(\ref{pol_Gamma0_T0}) describes vacuum contribution and takes into account
only interband processes while the second one represents intraband and interband
contributions when the chemical potential lies in the conduction or valence band.
Notice that this second term does not receive contribution from the
terms with $n=n'$, $\lambda=+1$.

In the gapless case ($\Delta=0$) we have
\be
\label{pol_Gamma0_T0_Delta0}
\Pi(\Omega,\bq)=\frac{e^2N_f}{4\pi l^2}\sum_{n,n'=0}^{\infty}\sum_{\zeta=\pm}
\frac{Q_{nn'}^{-+}(y,0)}{M_n+M_{n'}+\zeta(\Omega+i0)}
+\frac{e^2N_f}{4\pi l^2}\sum_{n=0}^{\infty}\sum_{n'=1}^{N_F}
\sum_{\lambda,\zeta=\pm}\frac{Q_{nn'}^{\lambda+}(y,0)}
{\lambda M_n-M_{n'}+\zeta(\Omega+i0)}\,.
\ee
Expressions~(\ref{pol_Gamma0_T0}), (\ref{pol_Gamma0_T0_Delta0}),
 coincide with the polarization function calculated
in~\cite{Shizuya2007PRB}. Refs.~\cite{Roldan} considered only gapless case
and obtained the expression similar to Eq.(\ref{pol_Gamma0_T0_Delta0}) but with twice larger contribution
of the lowest Landau level ($n=0$), while the results of the
papers~\cite{Tahir2008JPCM,Berman2008PRB} are completely different from ours.

The static clean limit of the polarization function essentially depends on the
order of taking limits $\Omega\to0$ and $\Gamma_n\to0$. Indeed, taking
first the limit $\Omega=0$, the expression for the polarization
function~(\ref{pol_main}) reduces to
\be
\label{pol_static}
\begin{split}
\Pi(0,\bq)=&\frac{e^2N_f}{8\pi^3l^2T}
\sum_{n=0}^{n_c}\sum_{\lambda=\pm}Q_{nn}^{\lambda\lambda}(y,\Delta)
\re\psi'\biggl(\frac12+\frac{\mu-\lambda M_n+i\Gamma_n}{2i\pi T}\biggr)\\
&+\frac{e^2N_f}{4\pi^2l^2}\sum_{\substack{n,n'=0\\\lambda n\,\ne\lambda'\!n'}}^{n_c}
\sum_{\lambda,\lambda'=\pm}Q_{nn'}^{\lambda\lambda'}(y,\Delta)
\im\frac{\psi\bigl(\frac12+\frac{\mu-\lambda M_n+i\Gamma_n}{2i\pi T}\bigr)
-\psi\bigl(\frac12+\frac{\mu-\lambda'M_{n'}+i\Gamma_{n'}}{2i\pi T}\bigr)}
{\lambda M_n-\lambda'M_{n'}-i(\Gamma_n-\Gamma_{n'})}\,,
\end{split}
\ee
where we took into account that the numerator of the third term in square brackets
in Eq.(\ref{pol_main}) vanishes at $\Omega=0$.
Here we also introduced the ultraviolet cutoff $n_{c}$ due to the divergence
of the sum over the Landau levels at finite width $\Gamma_{n}$. This cutoff is estimated to be
$n_c\sim10^4/B[T]$ due to finiteness of the bandwidth~\cite{Roldan}. The expression
(\ref{pol_static}) for static polarization  is obviously a real function.
In the clean graphene limit, $\Gamma_{n}=0$,  we get
\be
\label{pol_static_clean}
\Pi(0,\bq)=\frac{e^2N_f}{16\pi l^2T}
\sum_{n=0}^{\infty}\sum_{\lambda=\pm}\frac{Q_{nn}^{\lambda\lambda}(y,\Delta)}
{\cosh^2\bigl(\frac{\mu-\lambda M_n}{2T}\bigr)}
-\frac{e^2N_f}{4\pi l^2}\sum_{\substack{n,n'=0\\\lambda n\,\ne\lambda'\!n'}}^{\infty}
\sum_{\lambda,\lambda'=\pm}Q_{nn'}^{\lambda\lambda'}(y,\Delta)
\frac{n_F(\lambda M_n)-n_F(\lambda'M_{n'})}{\lambda M_n-\lambda'M_{n'}}\,
\ee
(the sum over the Landau levels is convergent).
On the other hand, if we take limit $\Omega\to0$ in~(\ref{pol_clean}), i.e., after
setting $\Gamma_n=0$, we obtain the expression~(\ref{pol_static_clean}) without the first term.
This term gives the contribution from the intra-level transitions ($n\leftrightarrow n$)
even at zero width of Landau levels.
At zero temperature it turns into the sequence of delta-functions $\delta(\mu\pm M_n)$
and does not contribute at integer filling factors of Landau levels ($\nu=0,\pm2,\pm6,\pm10,\ldots$).
Therefore, for $T=0$, we arrive at the same expression~(\ref{pol_Gamma0_T0}) with $\Omega=0$.

\section{Static screening}
\label{SecIV}

The screening of the static Coulomb potential $\phi_0(r)=Ze/r$ is determined by the static
polarization function,
\be
\label{screened_potential}
\phi(r)=\frac{Z e}{\e_0}\int\frac{d^2q}{2\pi}\frac{\exp(i\bq\br)}{q+({2\pi}/{\e_0})\Pi(0,\bq)}
=\frac{Ze}{\e_0}\int\limits_{0}^{\infty}\frac{dq\,q J_{0}(q r)}{q+({2\pi}/{\e_0})\Pi(0,q)}\,,
\ee
where $J_{0}(z)$ is the Bessel function
and $\e_0$ is the background dielectric constant due to the substrate.

In what follows we assume that even in the case of clean graphene, the limit $\Gamma_n\to0$
of the polarization function is taken after the limit $\Omega\to0$
when calculating the screened potential~(\ref{screened_potential}). This order of limits,
which leads to the expression~(\ref{pol_static_clean}) for $\Pi(0,\bq)$, seems to be more
natural due to the fact that real graphene samples can not be
completely free from impurities and some broadening of the Landau levels always occurs.

In general case, the static polarization function
is given by the expression ~(\ref{pol_static}) which does not have singularities
(like, for example, the discontinuity of the second derivative at $q=2\mu/v_F$
in the absence of magnetic field~\cite{Wunsch2006NJP}).
Therefore, the asymptotical behavior of the screened potential at small or large distances
is determined solely by the asymptotics of $\Pi(0,\bq)$ at large or small wavevectors, respectively.
At large momenta we have the zero magnetic field result,
\be
\Pi(0,\bq)\simeq\frac{e^2N_f|\bq|}{8v_F}\,,
\quad
\bq\to\infty\,,
\ee
and~(\ref{screened_potential}) implies
\be
\label{pol-asympt-short}
\phi(r)\simeq\frac{Ze}{\e_0^*r}\,,
\quad
r\to0\,,
\ee
where
\be
\e_0^*=\e_0+\frac{\pi e^2N_f}{4v_F}\approx\e_0+3.4
\ee
is the ``effective'' background dielectric constant ($N_{f}=2$).
At small values of a wavevector ($\bq\to0$), the static polarization function~(\ref{pol_static}) behaves as
\be
\label{pol-asympt-long}
\Pi(0,\bq)\simeq\frac{\e_0}{2\pi}\left(q_F+a\bq^{2}\right)\,,
\quad
\bq\to0\,.
\ee
In the case $q_F\ne0$, we find from~(\ref{screened_potential}) the following asymptotical behavior
\be
\phi(r) \simeq\frac{Ze}{\epsilon_0q_F^2r^3}\,,\quad r\to\infty\,,
\label{phi-asympt-short}
\ee
that describes the Thomas-Fermi screening in graphene~\cite{Wunsch2006NJP}.
In contrast to the three-dimensional case where for nonzero charge density the Coulomb
potential $1/r$ is replaced by an exponential decreasing one, in two-dimensional case we have
$1/r^{3}$ behavior at large $r$, which is the well known fact~\cite{Ando1982RMP}.
The strength of the screening is determined by the magnitude of $q_F=(2\pi/\e_0)\Pi(0,0)$.

At zero momentum $\bq=0$ only the first term in Eq.~(\ref{pol_static}) contributes and we get
\be
\label{pol_Omega0_p0}
\Pi(0,0)=\frac{e^2N_f}{4\pi^3l^2T}\sum_{n=0}^{n_c}\sum_{\lambda=\pm}(2-\delta_{0n})
\re\psi'\biggl(\frac12+\frac{\mu-\lambda M_n+i\Gamma_{n}}{2i\pi T}\biggr)\,.
\ee
The above polarization function obeys the following relation~\cite{Davies}
\be
\label{relation_pol_density}
\Pi(0,0)=e^2\frac{\partial}{\partial\mu}\rho(\mu,T)=e^2\int\limits_{-\infty}^\infty
\frac{d\epsilon\,D(\epsilon)}{4T\cosh^2\bigl(\frac{\epsilon-\mu}{2T}\bigr)}\,,
\ee
where $D(\epsilon)$ is the density of states in graphene with impurities
in magnetic field~\cite{Sharapov2004PRB},
\be
D(\epsilon)=\frac{N_f}{2\pi^2l^2}\sum_{n=0}^{n_c}
\sum_{\lambda=\pm}\frac{(2-\delta_{0n})\Gamma_n}{(\epsilon-\lambda M_n)^2+\Gamma_n^2}\,,
\ee
and $\rho(\mu,T)$ is the density of Dirac quasiparticles,
\be
\rho(\mu,T)=\int\limits_{-\infty}^\infty d\epsilon\,D(\epsilon)\bigl[n_F(\epsilon)-\theta(-\epsilon)\bigr]\,.
\ee
At zero temperature and finite scattering rate the quantity $\Pi(0,0)$ is proportional to
the density of states at the Fermi surface,
\be
\Pi(0,0)=\frac{e^2N_f}{2\pi^2l^2}\sum_{n=0}^{n_c}
\sum_{\lambda=\pm}\frac{(2-\delta_{0n})\Gamma_n}{(\mu-\lambda M_n)^2+\Gamma_n^2}
=e^2D(\mu)\,.
\ee
It is an oscillating function of chemical potential and a magnetic
field~\cite{Sharapov2004PRB}, and therefore,
the screened potential at large distances oscillates with changing $\mu$ at a fixed magnetic field,
or with changing $B$ at fixed $\mu$.

\begin{figure}
\includegraphics[width=7cm]{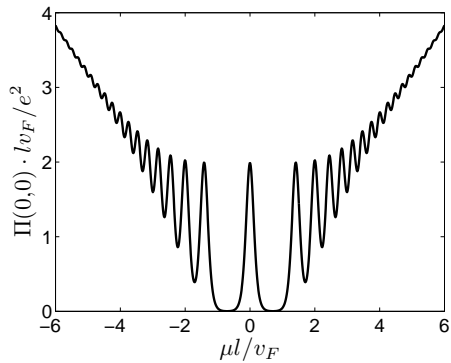}
\caption{Long wavelength limit of the static polarization function $\Pi(0,0)$ at
$\Gamma=0$, $\Delta=0$, $T=0.08v_F/l$.}
\label{oscillations}
\end{figure}
For $\Gamma_{n}=0$ and finite temperature, (\ref{pol_Omega0_p0}) reduces to the expression
\be
\label{pol_Omega0_p0_Gamma0}
\Pi(0,0)=-\frac{e^2N_f}{8\pi l^2T}\sum_{n=0}^{n_c}\sum_{\lambda=\pm}
\frac{2-\delta_{0n}}{\cosh^2\bigl(\frac{\mu-\lambda M_n}{2T}\bigr)}\,,
\ee
which has qualitatively the similar oscillatory behavior, see Fig.~\ref{oscillations}.
The weak magnetic field limit ($l\to\infty$) of the above expression can be
obtained by replacing $n\to k^2l^2/2$, with the sum turning into the integral
over $k$, resulting in
\be
\Pi(0,0)=\frac{e^2N_fT}{\pi v_F^2}\biggl[\ln\biggl(2\cosh\Bigl(
\frac{\Delta+\mu}{2T}\Bigr)\biggr)
-\frac\Delta{2T}\tanh\Bigl(\frac{\Delta+\mu}{2T}\Bigr)+(\mu\to-\mu)\biggr]\,,
\ee
which agrees with~\cite{Gorbar2002PRB}.

\begin{figure}
\includegraphics[width=7cm]{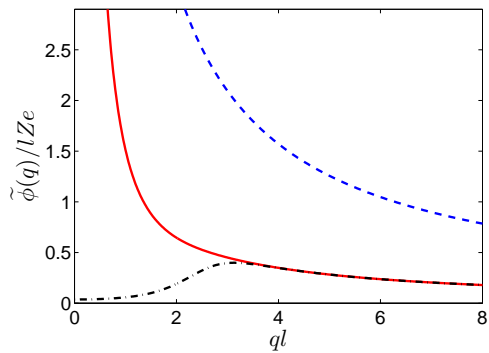}
\caption{Fourier transform of the screened Coulomb potential at $\Gamma=0$, $\Delta=0$, $T=0.01v_F/l$.
Dot-dashed (black) line: $\mu=0.01v_F/l$, solid (red) line:  $\mu=0.1v_F/l$. Dashed
(blue) line shows the unscreened case.}
\label{qpotential}
\end{figure}

\begin{figure}
\includegraphics[width=7cm]{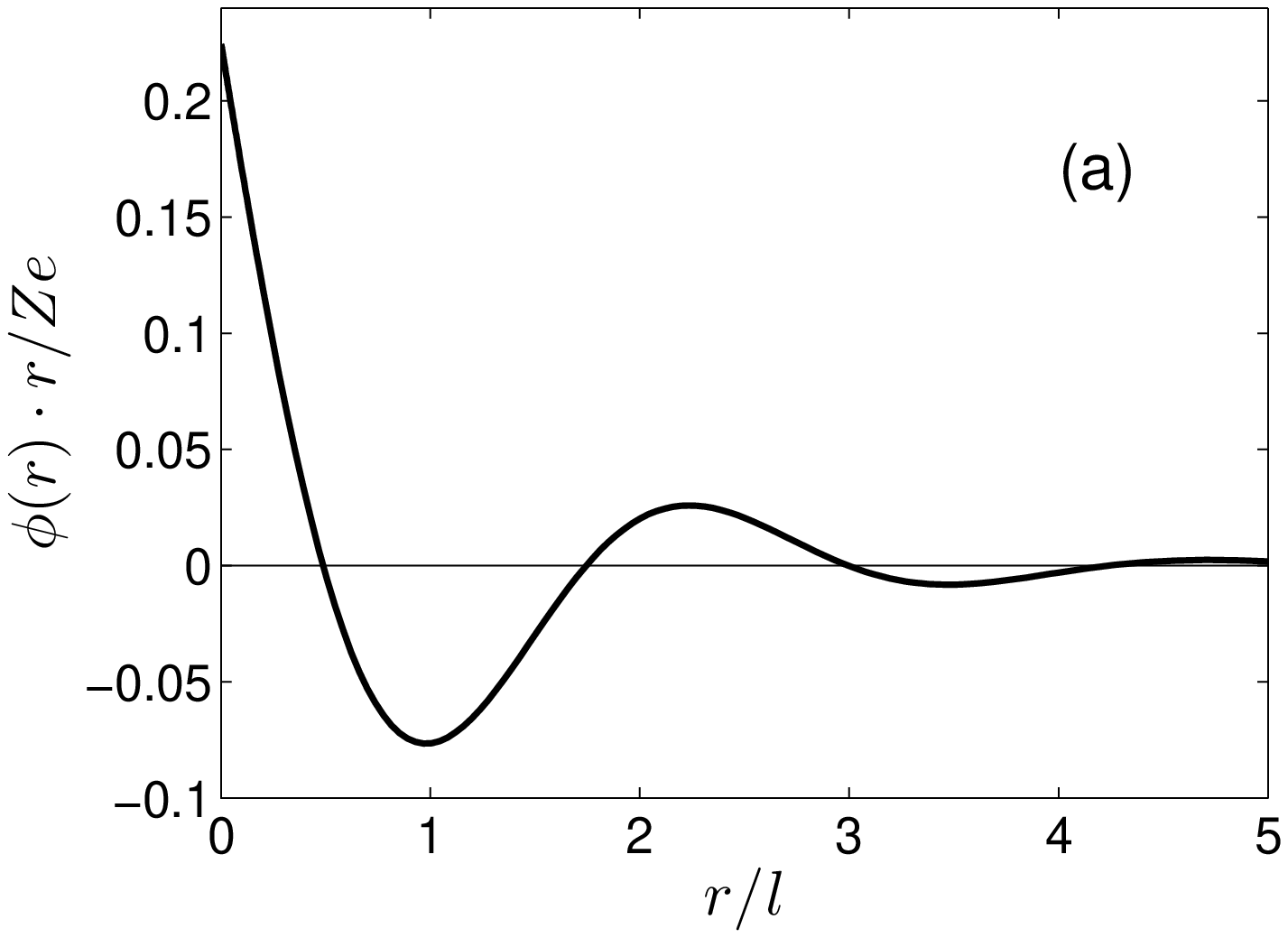}
\includegraphics[width=7cm]{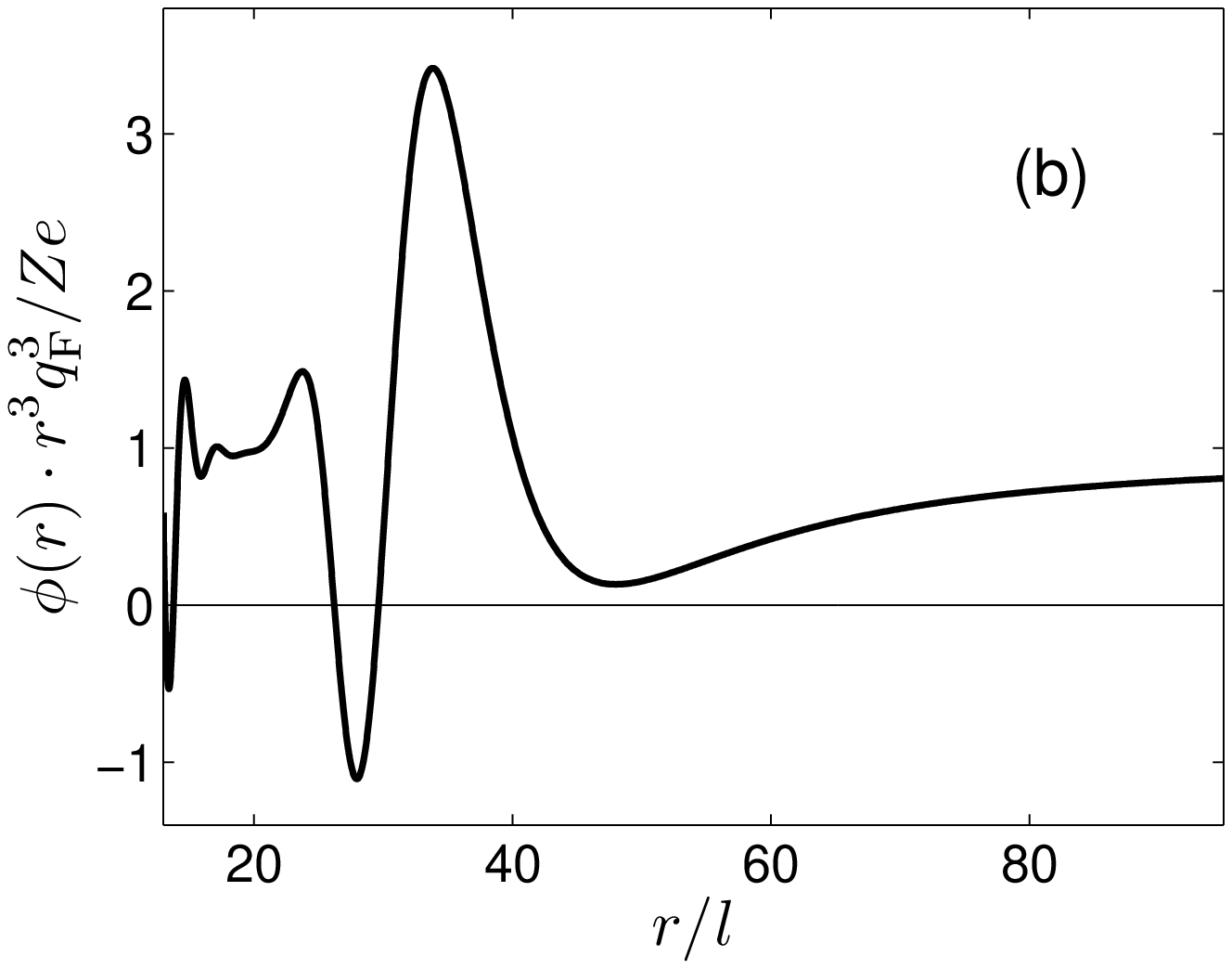}
\includegraphics[width=7cm]{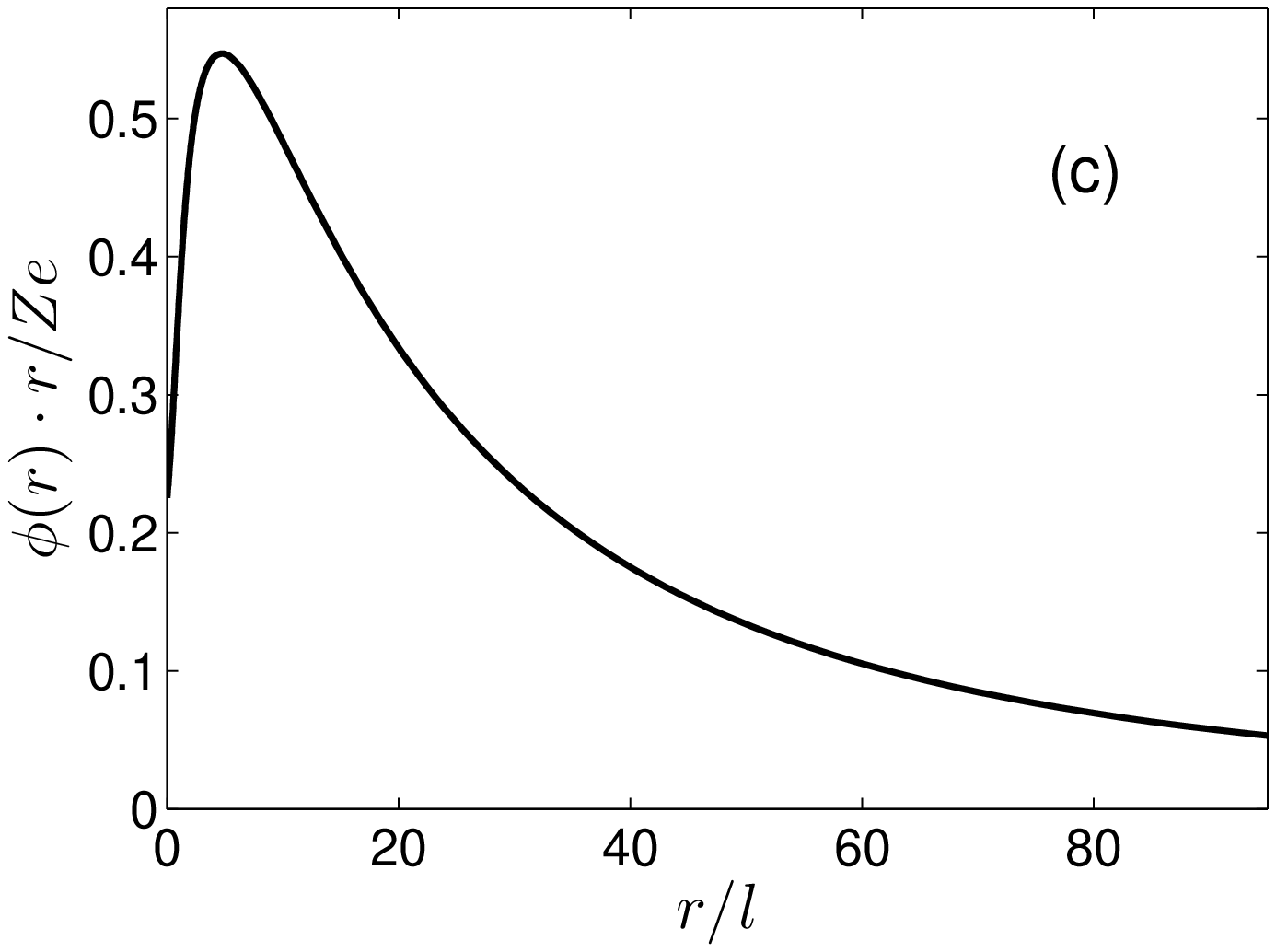}
\includegraphics[width=7cm]{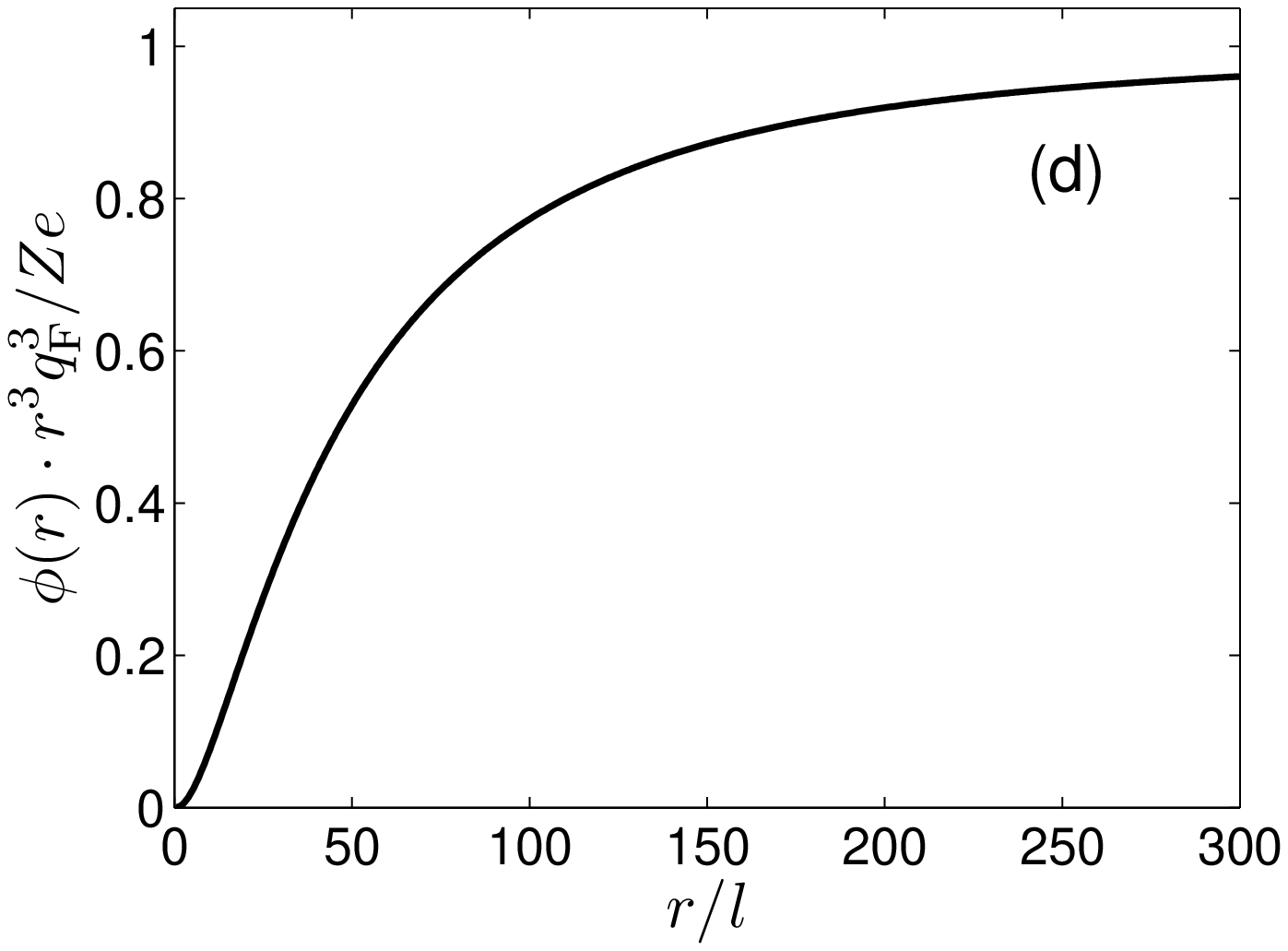}
\caption{Screened Coulomb potential at small ((a),(c))
and large ((b),(d)) distances. Here $\Gamma=0$, $\Delta=0$, $T=0.01v_F/l$, and
the value of the chemical potential is $\mu=0.01v_F/l$ at (a),(b) and $\mu=0.1v_F/l$ at (c),(d).}
\label{potential}
\end{figure}

Some numerical results for the screened Coulomb potential in the case of clean gapless graphene
at finite temperature are shown in Figs.~\ref{qpotential},~\ref{potential} (we used $\e_0=1$).
The figure~\ref{qpotential} shows the Fourier transform of the potential~(\ref{screened_potential}),
\be
\widetilde\phi(q)=\frac{2\pi Ze}{\e_0q+2\pi\Pi(0,q)}\,,
\ee
and the figure~\ref{potential} represents the screened potential~(\ref{screened_potential}) itself.
While the asymptotics of $\phi(r)$ is always given by Eqs.(\ref{pol-asympt-short}),
(\ref{phi-asympt-short}), its behavior at intermediate
distances can be qualitatively different,
depending on the values of the parameters $lT$ and $l\mu$.
If the temperature is sufficiently low ($T\lesssim0.1v_F/l$)
and the chemical potential lies in the vicinity
of one of the Landau levels, the coefficient $a$ in~(\ref{pol-asympt-long}) is negative
and $1/q_F\ll l\ll|a|$.
In this case the screened potential~(\ref{screened_potential}) oscillates at intermediate distances
$1/q_F<r<|a|$, as shown in Fig.~\ref{potential}(a),(b).
When the chemical potential lies away from the Landau levels, or the temperature is
larger than $0.4v_F/l$, the coefficient
$a$ is positive and $\phi(r)$ does not oscillate (Fig.~\ref{potential}(c),(d)).
In this case the asymptotic behavior of the screened potential for $r\gg l$ is given by
\ba
\label{screening_asympt_1}
\phi(r)\simeq\frac{Ze}{\e_0}\int\frac{d^2q}{2\pi}\frac{\exp(i\bq\br)}{q+q_F+a q^{2}}
=\frac{\pi Ze}{2\e_0 a(q_{1}-q_{2})}\Bigl\{q_{1}\left[H_{0}(q_{1}r)-Y_{0}(q_{1}r)\right]
-q_{2}\bigl[H_{0}(q_{2}r)-Y_{0}(q_{2}r)\bigr]\Bigr\}\,,
\ea
where $q_{1,2}=(1\pm\sqrt{1-4a q_{F}})/2a$, $H_{0}(z)$ is the Struve function,
and $Y_{0}(z)$ is the Bessel functions of the second kind.

Now let us consider the case when both temperature and scattering rate are zero. In this case
\be
\Pi(0,0)=\frac{e^2N_f}{2\pi l^2}\sum_{n=0}^{n_c}\sum_{\lambda=\pm}
(2-\delta_{0n})\delta(\mu-\lambda M_n)
=e^2D_0(\mu)\,,
\ee
where $D_0(\mu)$ is the DOS at the Fermi surface for the clean graphene~\cite{Sharapov2004PRB}.
When the Fermi level lies between Landau levels (which corresponds to the
integer fillings) the above expression vanishes, i.e., $q_F=0$.
Restricting ourselves to these integer fillings and setting
$\Omega=0$ in~(\ref{pol_Gamma0_T0}) or, equivalently, setting $T=0$, $\Gamma_n=0$
in~(\ref{pol_static}), we obtain
\be
\Pi(0,\bq)=\frac{e^2N_f}{2\pi l^2}\sum_{n,n'=0}^{n_c}
\frac{Q_{nn'}^{-+}(y,\Delta)}{M_n+M_{n'}}
-\frac{e^2N_f}{2\pi l^2}\theta(\mu^2-\Delta^2)\sum_{n'=0}^{n_c}\sum_{\substack{n=0\\n\ne\lambda'n'}}^{N_F}
\sum_{\lambda'=\pm}\frac{Q_{nn'}^{\lambda'+}(y,\Delta)}{M_n-\lambda'M_{n'}}\,.
\ee
Now the transitions between levels $n\leftrightarrow\pm n\pm1$ give the main contribution
at long wavelengths, because of the following asymptotics of the functions~(\ref{Q}) at $y\to0$ :
\ba
Q_{n,n+1}^{\lambda\lambda'}(y,\Delta)&=&Q_{n+1,n}^{\lambda\lambda'}(y,\Delta)=
y\biggl[2n+1+\lambda\lambda'\biggl(\frac{nM_{n+1}}{M_n}+\frac{(n+1)M_n}{M_{n+1}}\biggr)\biggr]
+\mathcal O(y^4)\,,\qquad n\ge0\,,\\
Q_{nn'}^{\lambda\lambda'}(y,\Delta)&=&\mathcal O(y^4)\,,\qquad n\ne n\pm1\,,\quad\lambda n\ne\lambda'n'\,.
\ea
This leads to the behavior
\be
\Pi(0,\bq)\simeq \frac{\e_0}{2\pi}a\bq^2\,,\qquad
|\bq|\ll1/l\,.
\ee
The coefficient $a$ at zero temperature and scattering rate is always positive and
depends on the number of filled Landau levels $N_F$
and the gap $\Delta$. It is evaluated to be
\be
\label{a}
a(N_F,\Delta)=\frac{e^2N_fl}{\sqrt2\e_0v_F}
\biggl(F(d)+\theta(\mu^2-\Delta^2)\sum_{n=0}^{N_F}
\frac{(2-\delta_{0n})(3n+2d)}{\sqrt{n+d}}\biggr)\,,
\ee
where $d=l^2\Delta^2/2v_F^2$ is the dimensionless gap parameter, and we define the function $F(d)$ as
\be
F(d)=\sum_{n=1}^{n_c}\bigl(\sqrt{n+d}-\sqrt{n-1+d}\bigr)^3
\biggl(1+\frac{d}{\sqrt{n+d}\sqrt{n-1+d}}\biggr)\,.
\ee
At zero gap and $N_F=0$, we obtain,
in agreement with~\cite{Shpagin1996,Gorbar2002PRB,Shizuya2007PRB},
\be
a(0,0)=\frac{e^2N_fl}{\sqrt2\e_0v_F}F(0)\,,\qquad
F(0)=-6\zeta(-1/2)-\frac1{4\sqrt{n_c}}+\mathcal O(n_c^{-3/2})\simeq1.247\,,
\ee
where $\zeta(z)$ is the Riemann zeta function.

From Eq.(\ref{screened_potential}) we obtain that at long distances the screening is absent
(the correction to the bare Coulomb potential is of the smaller order),
\be
\phi(r)\simeq\frac{Ze}{\e_0r}\biggl(1-\frac{a^2}{r^2}\biggr)\,,\qquad
r\gg l\,.
\ee

\section{Other limiting cases}
\label{SecV}

At zero momentum, only the terms with $\lambda n=\lambda' n'$ survive in~(\ref{pol_main}),
and the general expression for the polarization function simplifies to
\be
\label{pol_p0}
\Pi(\Omega,0)=\frac{ie^2N_f}{\pi l^2\Omega}\sum_{n=0}^{n_c}
\frac{(2-\delta_{0n})\Gamma_n}{\Omega+2i\Gamma_n}
\sum_{\lambda=\pm}Z^{\lambda\lambda}_{nn}(\Omega,\Gamma,\mu,T)+(\mu\to-\mu)\,.
\ee
At zero scattering rate and finite $\Omega$ the above expression vanishes,
while its limit $\Omega\to0$ at finite $\Gamma_n$ is given by the equation~(\ref{pol_Omega0_p0}).
Therefore, the static long-wavelength polarization function $\Pi(0,0)$ does not depend on the order
of taking limits $\Omega\to0$ and $\bq\to0$, unlike in the absence of magnetic field.

The strong magnetic field limit ($l\to0$) of the polarization function~(\ref{pol_main})
also depends on the ratio between the scattering rate and the frequency.
For $\Gamma_n/\Omega\ne0$ the main contribution comes only
from the lowest Landau level ($n=0$) and is given by the expression
\be
\Pi(\Omega,\bq)\simeq-\frac{e^2N_f}{2\pi^2l^2}\frac{\Gamma_0}{\Omega(\Omega+2i\Gamma_0)}\sum_{\lambda,\lambda'=\pm}
\biggl\{\psi\biggl(\frac12
+\frac{\lambda\mu+\lambda'\Delta+\Omega+i\Gamma_0}{2i\pi T}\biggr)
-\psi\biggl(\frac12
+\frac{\lambda\mu+\lambda'\Delta+i\Gamma_0}{2i\pi T}\biggr)\biggr\}\,.
\ee
However, this contribution vanishes in the clean graphene limit
(more exactly, for $\Gamma_0=0$ and nonzero $\Omega$).
In this case the transitions $n\leftrightarrow-n\pm1$ in~(\ref{pol_main}) dominate at
high magnetic field, resulting in
\be
\Pi(\Omega,\bq)\simeq\frac{\e_0}{2\pi}a(0,0)\bq^2\,,
\ee
which is equivalent to the static long wavelength limit of the polarization function
for the clean gapless graphene at zero temperature in the case when only the lowest
Landau level is filled.

\section{Summary}
\label{SecVI}

In this paper we have derived the exact analytical expression for the one-loop dynamical
polarization function in graphene, as a function of wavevector and frequency,
at finite chemical potential, temperature, band gap, and taking into account the finite
scattering rate of Dirac quasiparticles due to the presence of impurities.
The most general result is given in terms of the digamma function and generalized
Laguerre polynomials and has the form of double sum over Landau levels, Eq.(\ref{pol_main}).
In the clean graphene at zero temperature, for the integer fillings of Landau levels,
this function correctly reproduces the previously obtained results.
The derived expression for dynamical polarization can be used to calculate the dispersion
relation and the decay rate of magnetoplasmons depending on temperature and impurity rate.

The long-range behavior of the screened static Coulomb potential in graphene in magnetic field
is found to be essentially affected by the presence of impurities or the finite temperature.
When either the scattering rate or the temperature is nonzero, the usual Thomas-Fermi screening is present,
and the resulting potential decays as $\sim1/r^3$, which is typical for two-dimensional systems.
The strength of the screening oscillates as a function of chemical potential or a magnetic field.
If both scattering rate and temperature are zero, these oscillations turn into the sequence of delta-functions,
and for the integer fillings the screening is absent.

\section{Acknowledgments}
We are grateful to E.V. Gorbar, V.A. Miransky, R.~Rold\'an and I.A. Shovkovy for useful
discussions.
The work of V.P.G. was supported partially by the SCOPES grant
No. IZ73Z0\verb|_|128026 of Swiss NSF, by the grant SIMTECH No. 246937 of the European FP7 program,
by the joint grant RFFR-DFFD No. F28.2/083 of the Russian Foundation
for Fundamental Research (RFFR) and of the Ukrainian State Foundation for Fundamental Research (DFFD),
and by the Program of Fundamental Research of the Physics and Astronomy Division of
the NAS of Ukraine.

\appendix

\section{Calculation of polarization function}

After evaluation of the trace, the equation~(\ref{pol_init}) can be written in the following form
\be
\label{pol}
\Pi(i\Omega_s,\bq)=-\frac{e^2TN_f}{8\pi^2l^4}
\sum_{n,n'=0}^{n_c}\sum_{\lambda,\lambda'=\pm}\sum_{m=-\infty}^\infty
\frac{\Bigl(1+\frac{\lambda\lambda'\Delta^2}{M_nM_{n'}}\Bigr)\bigl[I_{nn'}^0(y)+I_{n-1,n'-1}^0(y)\bigr]
+\frac{4\lambda\lambda'v_F^2}{l^2M_nM_{n'}}I_{n-1,n'-1}^1(y)}
{(i\omega_m+\mu+i\Gamma_n\sgn\omega_m-\lambda M_n)
(i\omega_{m-s}+\mu+i\Gamma_{n'}\sgn\omega_{m-s}-\lambda'M_{n'})}\,,
\ee
where $y=\bq^2l^2/2$, and
\be
\label{I}
I_{nn'}^\alpha(y)=\int d^2r\,e^{-i\bq\br}\Bigl(\frac{\br^2}{2l^2}\Bigr)^\alpha\exp\Bigl(-\frac{\br^2}{2l^2}\Bigr)
L_n^\alpha\Bigl(\frac{\br^2}{2l^2}\Bigr)L_{n'}^\alpha\Bigl(\frac{\br^2}{2l^2}\Bigr)\,,\qquad
\alpha=0,1.
\ee
The above expression is nonzero only for $n,n'\ge0$.
Integrating over the angle and making the change of the variable $\br^2=2l^2t$, we get
\ba
\label{I_1}
I_{nn'}^\alpha(y)&=&2\pi l^2\int_0^\infty dt\,e^{-t}t^\alpha J_0\bigl(2\sqrt{yt}\bigr)
L_n^\alpha(t)L_{n'}^\alpha(t)\nonumber\\
&=&2\pi l^2(-n'-1)^{\alpha}\int_0^\infty dt\,e^{-t}J_0\bigl(2\sqrt{yt}\bigr)
L_n^\alpha(t)L_{n'+\alpha}^{-\alpha}(t)\,,\qquad
\alpha=0,1,
\ea
where we have used
\be
\label{Laguerre}
L_l^k(x)=(-x)^{-k}\frac{(l+k)!}{l!}L_{l+k}^{-k}(x)\,,\qquad
l\ge0\,,\quad k+l\ge0\,.
\ee
Now, using the formula 7.422.2 in~\cite{Gradshteyn}
\be
\int\limits_0^\infty dx\,x^{\nu+1}e^{-\alpha x^2}J_{\nu}(bx)L_m^{\nu-\sigma}(\alpha x^2)
L_n^\sigma(\alpha x^2)=
(-1)^{m+n}(2\alpha)^{-\nu-1}b^\nu
e^{-\frac{b^2}{4\alpha}}L_m^{\sigma-m+n}\biggl(\frac{b^2}{4\alpha}\biggr)
L_n^{\nu-\sigma+m-n}\biggl(\frac{b^2}{4\alpha}\biggr)\,,
\ee
we obtain from~(\ref{I_1})
\ba
\label{I_result}
I_{nn'}^\alpha(y)&=&2\pi l^2(-1)^{n-n'}(n'+1)^\alpha e^{-y}L_n^{n'-n}(y)L_{n'+\alpha}^{n-n'}(y)\nonumber\\
&=&2\pi l^2\frac{(n_<+\alpha)!}{n_>!}e^{-y}y^{|n-n'|}L_{n_<}^{|n-n'|}(y)L_{n_<+\alpha}^{|n-n'|}(y)\,,\qquad
\alpha=0,1,
\ea
where we again used the formula~(\ref{Laguerre}) and the symmetry
$I^\alpha_{nn'}(y)=I^\alpha_{n'n}(y)$ which follows from~(\ref{I}).
Now we can rewrite~(\ref{pol}) as
\be
\label{Pi}
\Pi(i\Omega_s,\bq)=-\frac{e^2TN_f}{4\pi l^2}
\sum_{n,n'=0}^{n_c}\sum_{\lambda,\lambda'=\pm}
Q_{nn'}^{\lambda\lambda'}(y,\Delta)\,\mathcal I\,,
\ee
where the functions $Q_{nn'}^{\lambda\lambda'}(y,\Delta)$ are defined in~(\ref{Q}) and
\be
\mathcal I=\sum_{m=-\infty}^\infty\frac1{(i\omega_m+\mu+i\Gamma_n\sgn\omega_m-\lambda M_n)
(i\omega_{m-s}+\mu+i\Gamma_{n'}\sgn\omega_{m-s}-\lambda'M_{n'})}\,.
\ee
To evaluate this sum, we expand it in terms of partial fractions and split into four sums in the following way:
\be
\begin{split}
&\mathcal I=\frac1{\lambda M_n-\lambda'M_{n'}-i\Omega_s-i(\Gamma_n-\Gamma_{n'})}
\sum_{m=s}^\infty\biggl(\frac1{i\omega_m+\mu+i\Gamma_n-\lambda M_n}-
\frac1{i\omega_{m-s}+\mu+i\Gamma_{n'}-\lambda'M_{n'}}\biggr)\\
&+\frac1{\lambda M_n-\lambda'M_{n'}-i\Omega_s-i(\Gamma_{n'}-\Gamma_n)}
\sum_{m=-\infty}^{-1}\biggl(\frac1{i\omega_m+\mu-i\Gamma_n-\lambda M_n}-
\frac1{i\omega_{m-s}+\mu-i\Gamma_{n'}-\lambda'M_{n'}}\biggr)\\
&+\frac1{\lambda M_n-\lambda'M_{n'}-i\Omega_s-i(\Gamma_n+\Gamma_{n'})}
\biggl(\sum_{m=0}^\infty-\sum_{m=s}^\infty\biggr)\biggl(\frac1{i\omega_m+\mu+i\Gamma_n-\lambda M_n}-
\frac1{i\omega_{m-s}+\mu-i\Gamma_{n'}-\lambda'M_{n'}}\biggr)\,.
\end{split}
\ee
Now, making the change $m\to m+s$ in the first and the last sums, and $m\to-m-1$ in the second one,
and using the summation formula
\be
\sum_{n=0}^\infty\biggl(\frac1{n+a}-\frac1{n+b}\biggr)=\psi(b)-\psi(a)\,,
\ee
we obtain
\be
\begin{split}
-T\,\mathcal I={}&\frac{Z_{nn'}^{\lambda\lambda'}(i\Omega_s,\Gamma,\mu,T)}
{\lambda M_n-\lambda'M_{n'}-i\Omega_s-i(\Gamma_n-\Gamma_{n'})}
+\frac{Z_{n'n}^{-\lambda',-\lambda}(i\Omega_s,\Gamma,-\mu,T)}
{\lambda M_n-\lambda'M_{n'}-i\Omega_s-i(\Gamma_{n'}-\Gamma_n)}\\
&-\frac{Z_{nn}^{\lambda\lambda}(i\Omega_s,\Gamma,\mu,T)
+Z_{n'n'}^{\lambda'\lambda'}(-i\Omega_s,-\Gamma,\mu,T)}
{\lambda M_n-\lambda'M_{n'}-i\Omega_s-i(\Gamma_n+\Gamma_{n'})}\,,
\end{split}
\ee
where the functions $Z_{nn'}^{\lambda\lambda'}(\Omega,\Gamma,\mu,T)$ are defined in~(\ref{Z}).
Using the above equation and the relation
\be
Z_{n'n'}^{\lambda'\lambda'}(-i\Omega_s,-\Gamma,\mu,T)
=Z_{n'n'}^{-\lambda',-\lambda'}(i\Omega_s,\Gamma,-\mu,T)\,,
\ee
which follows from the formula
\be
\label{digamma_property}
\psi(1-z)=\psi(z)+\pi\cot(\pi z)\,,
\ee
we can rewrite~(\ref{Pi}) as
\be
\begin{split}
\Pi(i\Omega_s,\bq)={}&\frac{e^2N_f}{4\pi l^2}
\sum_{n,n'=0}^{n_c}\sum_{\lambda,\lambda'=\pm}Q_{nn'}^{\lambda\lambda'}(y,\Delta)
\biggl[\frac{Z_{nn'}^{\lambda\lambda'}(i\Omega_s,\Gamma,\mu,T)}
{\lambda M_n-\lambda'M_{n'}-i\Omega_s-i(\Gamma_n-\Gamma_{n'})}\\
&+\frac{Z_{n'n}^{-\lambda',-\lambda}(i\Omega_s,\Gamma,-\mu,T)}
{\lambda M_n-\lambda'M_{n'}-i\Omega_s-i(\Gamma_{n'}-\Gamma_n)}
-\frac{Z_{nn}^{\lambda\lambda}(i\Omega_s,\Gamma,\mu,T)
+Z_{n'n'}^{-\lambda',-\lambda'}(i\Omega_s,\Gamma,-\mu,T)}
{\lambda M_n-\lambda'M_{n'}-i\Omega_s-i(\Gamma_n+\Gamma_{n'})}\biggr]\,.
\end{split}
\ee
Making the analytic continuation from Matsubara frequencies by replacing
$i\Omega_s\to\Omega+i0$,
we finally arrive at~(\ref{pol_main}). At constant scattering rate $\Gamma_n=\Gamma$ the result
simplifies to
\be
\begin{split}
\label{Pi-constGamma}
\Pi(\Omega,\bq)=\frac{e^2N_f}{4\pi l^2}
\sum_{n,n'=0}^{\infty}\sum_{\lambda,\lambda'=\pm}Q_{nn'}^{\lambda\lambda'}(y,\Delta)
\biggl[&\frac{Z_{nn'}^{\lambda\lambda'}(\Omega,\Gamma,\mu,T)+Z_{n'n}^{-\lambda',-\lambda}(\Omega,\Gamma,-\mu,T)}
{\lambda M_n-\lambda'M_{n'}-\Omega}\\
&-\frac{Z_{nn}^{\lambda\lambda}(\Omega,\Gamma,\mu,T)
+Z_{n'n'}^{-\lambda',-\lambda'}(\Omega,\Gamma,-\mu,T)}
{\lambda M_n-\lambda'M_{n'}-\Omega-2i\Gamma}\biggr]\,.
\end{split}
\ee
One can check that the first term in square brackets does not have poles at
$\Omega=\lambda M_n-\lambda'M_{n'}$ since the numerator vanishes at this point,
\ba
&&Z_{nn'}^{\lambda\lambda'}(\Omega,\Gamma,\mu,T)+Z_{n'n}^{-\lambda',-\lambda}(\Omega,\Gamma,-\mu,T)
=\nonumber\\
&&-\frac\epsilon{4\pi^2T}\biggl[\psi^{\prime}\biggl(\frac12+\frac{\mu-\lambda'M_{n'}
+i\Gamma}{2i\pi T}\biggr)+\psi^{\prime}\biggl(\frac12-\frac{\mu-\lambda M_n-i\Gamma}{2i\pi T}\biggr)\biggr],
\quad \Omega=\lambda M_n-\lambda'M_{n'}+\epsilon,\quad \epsilon\to0.
\label{stzero}
\ea
At $\Gamma\to0$ the denominators in~Eq.(\ref{Pi-constGamma}) become equal, and the overall numerator reads
\be
\label{splusr}
\begin{split}
Z_{nn'}^{\lambda\lambda'}&(\Omega,0,\mu,T)+Z_{n'n}^{-\lambda',-\lambda}(\Omega,0,-\mu,T)
- Z_{nn}^{\lambda\lambda}(\Omega,0,\mu,T)
-Z_{n'n'}^{-\lambda',-\lambda'}(\Omega,0,-\mu,T)\\
&=\frac1{2\pi i}\biggl\{-\biggl[\psi\biggl(\frac12-\frac{\mu-\lambda M_n}{2i\pi T}\biggr)
-\psi\biggl(\frac12+\frac{\mu-\lambda M_n}{2i\pi T}\biggr)\biggr]
+\biggl[\psi\biggl(\frac12-\frac{\mu-\lambda'M_{n'}}{2i\pi T}\biggr)-\psi\biggl(\frac12+
\frac{\mu-\lambda'M_{n'}}{2i\pi T}\biggr)\biggr]\biggr\}\\
&=n_F(\lambda'M_{n'})-n_F(\lambda M_n)\,,
\end{split}
\ee
where we used the property~(\ref{digamma_property}) of the digamma function.

\section{Schwinger proper-time calculation of polarization function in magnetic field}

The general expression (\ref{pol_main}) for the polarization function as a double sum
over the Landau levels is useful for high magnetic fields. Clearly, for weak fields
Eq.(\ref{pol_main}) is not convenient since we need to keep many terms in the double sum.
In general, when $\Gamma$ depends on the Landau index $n$ it is impossible even to get a closed
expression for the quasiparticle propagator, not to mention the polarization function itself.
In principle, it is possible to perform summation in Eq.(\ref{prop}) for $\Gamma=const$
and $\mu\neq0$ but the expression obtained looks rather cumbersome for further work with it.
Therefore, we consider in this section only the case $\Gamma=\mu=0$. Using the
identity ${1}/{a}=\int\limits_{0}^{\infty}dt\,e^{-a t}, a>0$ for introducing the
proper-time coordinate $t$, and the formula \cite{Erdelyi}
\be
\sum\limits_{n=0}^{\infty}L^{\alpha}_{n}(x)z^{n}=(1-z)^{-\alpha-1}\exp\frac{x z}{z-1},
\quad |z|<1,
\ee
we get a closed expression for the fermion propagator:
\ba
S(i\omega_m, \mathbf{r})&=&\frac{1}{4\pi i v_{F}^{2}}\int\limits_{0}^{\infty}dt
\,\exp\left[- t\frac{l^{2}(\omega_{m}^{2}+\Delta^{2})}{ v_{F}^{2}}-\frac{\mathbf{r}^{2}}
{4l^{2}}\coth t\right]\nonumber\\
&\times&\left\{(\gamma_{0}i\omega_{m}+\Delta)\bigl[P_{-}(1+\coth t)-P_{+}(1-\coth t)\bigr]
-i\frac{ v_{F}}{2l^{2}}\frac{\boldsymbol\gamma\mathbf{r}}{\sinh^{2}t}\right\}.
\label{propagator-repr-finiteT}
\ea
The integrals can be evaluated through confluent hypergeometric functions,
\ba
I_{1}(a,b)&=&\int\limits_{0}^{\infty}dt\,e^{-at-b\coth t}=\frac{1}{2}e^{-b}\Gamma
\left(\frac{a}{2}\right)\Psi\left(\frac{a}{2},0,2b\right),\quad I_{2}(a,b)
=\int\limits_{0}^{\infty}dt\,e^{-at-b\coth t}\coth t= -\frac{d I_{1}(a,b)}{db},
\nonumber\\
 I_{3}(a,b)
&=&\int\limits_{0}^{\infty}dt\,e^{-at-b\coth t}\coth^{2} t= \frac{d^{2} I_{1}(a,b)}{db^{2}},
\quad a=\frac{l^{2}(\omega_{m}^{2}+\Delta^{2})}{ v_{F}^{2}},\quad b=\frac{\mathbf{r}^{2}}
{4l^{2}}.
\ea
Hence, we have
\ba
S(i\omega_{m}; \mathbf{r})&=&\frac{e^{-\mathbf{r}^{2}/4l^{2}}}{4\pi iv_{F}^{2}}
\left\{(\gamma_{0}i\omega_{m}+\Delta)\left[P_{-}\Gamma\left(\frac{a}{2}\right)
\Psi\left(\frac{a}{2},1,\frac{\mathbf{r}^{2}}{2l^{2}}\right)+P_{+}\Gamma\left(1+\frac{a}{2}\right)
\Psi\left(1+\frac{a}{2},1,\frac{\mathbf{r}^{2}}{2l^{2}}\right)\right]\right.\nonumber\\
&+& \left. iv_{F}
\frac{\boldsymbol\gamma\mathbf{r}}{l^{2}}\Gamma\left(1+\frac{a}{2}\right)\Psi\left(1+
\frac{a}{2},2,\frac{\mathbf{r}^{2}}{2l^{2}}\right)\right\}.
\ea
Using the integral representation (\ref{propagator-repr-finiteT}) for the propagator,
we get from~(\ref{pol_init}) taking the trace and performing
the Gauss integration over coordinates,
\ba
\Pi(i\Omega_s,\mathbf{q})&=&-\frac{e^{2}T l^{2}N_{f}}{\pi v^{4}_{F}}\sum_{m=-\infty}^{\infty}
\int\limits_{0}^{\infty}\frac{d t\,d x}{\coth t+\coth x}\exp\left[-t\frac{l^2(\omega^{2}_{m}
+\Delta^{2})}{v^{2}_{F}}-x\frac{l^{2}(\omega_m'^{\,2}+\Delta^{2})}{v^{2}_{F}}
-\frac{\mathbf{q}^{2}l^{2}}{\coth t+\coth x}\right]\nonumber\\
&\times&\biggl[(\Delta^{2}-\omega_{m}\omega'_{m})(1+\coth t \coth x)
+\frac{v^{2}_{F}(\coth t+\coth x-\mathbf{q}^{2}l^{2})}{l^{2}\sinh^{2}(t+x)}\biggr],
\quad\omega'_{m} =\omega_{m}-\Omega_{s}.
\label{polarization-intermediate}
\ea
Introducing new variables, $t=z(1+v)/2$, $x=z(1-v)/2$, we obtain
\be
\begin{split}
\Pi(i\Omega_s,\mathbf{q})=&-\frac{T e^2N_f}{\pi l^2}\int\limits_0^\infty
{du}\int\limits_{-1}^1\frac{dv}2
\exp\biggl(-u\Delta^2-\frac{\cosh z-\cosh zv}{2\sinh z}
\mathbf{q}^2l^{2}\biggr)\\
&\times\biggl[\frac{z}{\sinh^{2}z}\biggl(1-\frac{\cosh z-\cosh zv}{2\sinh z}
\mathbf{q}^2l^{2}\biggr)
+{u\coth z}\biggl(\Delta^2+\frac{\Omega_s^2}{2}+\frac{\partial}{\partial u}-\frac{v}{u}
\frac{\partial}{\partial v}\biggr)
\biggr]R(u,v,\Omega_{s})\,,
\end{split}
\ee
where $u\equiv l^2z/v_F^2$ and the sum
\ba
\label{sum}
R(u,v,\Omega_m)=e^{-u(1-v^2)\Omega_s^2/4}
\sum_{m=-\infty}^\infty\exp\biggl[-4\pi^2T^2u\Bigl(m+\frac{1-s+sv}2\Bigr)^2\biggr]
\ea
can be written through the Jacobi elliptic function. For that we use the formula
\ba
\sum_{m=-\infty}^\infty q^{(m+c)^{2}}=q^{c^{2}}\theta_{3}\left(\frac{ic\ln q}{\pi},q\right)
=e^{i\pi c^{2}\tau}\theta_{3}(c\tau|\tau)=(-i\tau)^{-1/2}\theta_{3}(c|- 1/\tau),\quad q=e^{i\pi\tau},
\quad\im\tau>0,
\ea
where
\be
\theta_3(v,q)\equiv\theta_3(v|\tau)=1+2\sum_{n=1}^\infty q^{n^2}\cos(2\pi nv),\quad
\ee
and for the third equality we used the Jacobi imaginary  transformation. Hence the sum (\ref{sum})
takes the form
\be
\label{sumresult}
\begin{split}
R(u,v,\Omega_m)&=\frac{e^{-u(1-v^2)\Omega_s^2/4}}{2T\sqrt{\pi u}}
\theta_3\biggl[\frac12-\frac{(1-v)\Omega_m}{4\pi T}\,,e^{-1/(4uT^2)}\biggr]\\
&=\frac{e^{-u(1-v^2)\Omega_s^2/4}}{2T\sqrt{\pi u}}
\theta_4\biggl[\frac{(1+v)\Omega_m}{4\pi T}\,,e^{-1/(4uT^2)}\biggr].
\end{split}
\ee
Since
\be
\left(u\frac\partial{\partial u}-{v}\frac\partial{\partial v}\right)R(u,v,\Omega_s)=
\frac{e^{-u(1-v^2)\Omega_s^2/4}}{2T\sqrt{\pi u}}
\left(-\frac12-\frac{(1+v^{2})u\Omega_m^2}{4}
+u\frac\partial{\partial u}-{v}\frac\partial{\partial v}\right)\theta_4\left[
\frac{(1+v)\Omega_m}{4\pi T}\,,e^{-1/(4uT^2)}\right],
\ee
we write
\begin{eqnarray}
\Pi(i\Omega_s,\mathbf{q})&=&-\frac{ e^2N_f}{2\pi^{3/2}l^2}\int\limits_0^\infty
\frac{du}{\sqrt u}\int\limits_{-1}^1\frac{dv}2
\exp\left[-u\left(\Delta^2+\frac{(1-v^{2})\Omega_s^2}{4}\right)-\frac{\cosh z-\cosh zv}{2\sinh z}
\mathbf{q}^2l^{2}\right]\nonumber\\
&\times&\biggl\{\frac{z}{\sinh^{2}z}\left[1-\frac{\cosh z-\cosh zv}{2\sinh z}
\mathbf{q}^2l^{2}\right]
+{u\coth z}\left(\Delta^2+\frac{(1-v^{2})\Omega_s^2}{4}-\frac{1}{2u}+\frac{\partial}{\partial u}-\frac{v}{u}
\frac{\partial}{\partial v}\right)
\biggr\}\nonumber\\
&\times&\theta_4\left[
\frac{(1+v)\Omega_m}{4T}\,,e^{-1/(4uT^2)}\right].
\end{eqnarray}
The above integral is divergent at $u=0$ reflecting the primitive divergence of the polarization
function. Therefore, in order to get finite result one should regularize
the initial expression, for example, by subtracting the
same expression with $\Delta$ replaced by $M\to\infty$ (the Pauli-Villars regularization) which
means that we write
\be
\Pi(i\Omega_s,\mathbf{q})=\lim_{M\to\infty}\frac{-e^2N_f}{2\pi^{3/2}l^2}\int\limits_0^\infty
\frac{du}{\sqrt u}\int\limits_{-1}^1\frac{dv}2\Biggl\{\dots -(\Delta^2\to M^2)\Biggr\}.
\ee
Carefully separating the part with $M^{2}$ and taking into account that
\be
\lim_{M\to\infty}\int\limits_0^\infty\frac{d u}{\sqrt u}
\biggl[\exp(-u M^2)\left(\frac{1}{2u}+M^2\right)-\frac{1}{2u}\biggr]=0,
\ee
we finally get the following expression for the polarization function at finite temperature in a
magnetic field,
\begin{eqnarray}
&&\Pi(i\Omega_s,\mathbf{q})=-\frac{e^2N_f}{2\pi^{3/2}l^{2}}\int\limits_0^\infty\frac{du}{\sqrt u}
\int\limits_{-1}^1\frac{dv}2\Biggl\{\frac{\exp(-u\Delta^2)}{\sinh z}\biggl\{z\exp\left[-u
\left(\frac{1-v^2}4\Omega_s^2+\frac{\cosh z-\cosh zv}{2z\sinh z}\mathbf{q}^2v^{2}_{F}\right)\right]
\nonumber\\
&&\times\left[\frac1{\sinh z}\left(1-\frac{\cosh z-\cosh zv}{2\sinh z}\mathbf{q}^2l^{2}\right)+
\cosh z\left(\frac{2\Delta^2l^{2}}{v^{2}_{F}}+\frac{\Omega_s^2l^{2}}{2v^{2}_{F}}+
\frac{2}{\sinh 2z}+\mathbf{q}^2l^{2}\frac{\cosh z
\cosh z v-1}{2\sinh^2z}\right)\right]\nonumber\\
&&\times\theta_4\left[\frac{(1+v)\Omega_s}{4\pi T}\,,e^{-1/(4uT^2)}\right]
-\cosh z\theta_4\left[0\,,e^{-1/(4uT^2)}\right]\biggr\}-\frac{1}{z}
\Biggr\},
\label{polarization-finiteTB}
\end{eqnarray}
where we also performed the integration in parts of terms with derivatives over $u,v$.

Now we  consider several limiting cases of Eq.(\ref{polarization-finiteTB})
and compare them with expressions existing in the literature.
Taking the limit $T\to0$ is very easy since theta-functions turn into units. After some
transformations the zero temperature limit can be recast in the form
\be
\label{pol-Shpagin}
\begin{split}
\Pi(i\Omega_{s},\mathbf{q})={}&\frac{e^2N_f\mathbf{q}^2}{4\pi^{3/2}}\int\limits_0^\infty
\frac{du}{\sqrt u}\int\limits_{-1}^1\frac{dv}2\frac{z\cosh zv-zv\coth z\sinh zv}{\sinh z}\\
&\times\exp\left[-u\left(\Delta^2+\frac{1-v^2}4\Omega_{s}^2+\frac{\cosh z-\cosh zv}{2z\sinh z}
\mathbf{q}^2v^{2}_{F}\right)\right],
\end{split}
\ee
the result first obtained in Ref.\cite{Shpagin1996}.

On the other hand, taking the limit of zero field, $l\to\infty$, in Eq.(\ref{polarization-finiteTB}) we get
\ba
\Pi(i\Omega_{s},\mathbf{q})&=&-\frac{e^2N_f}{2\pi^{3/2}v^{2}_{F}}\int\limits_{0}^{\infty}
\frac{d u}{u^{3/2}}\int\limits_{-1}^1\frac{dv}2\left\{\exp\left[-u\left(\Delta^{2}
+\frac{1-v^{2}}{4}(\Omega^{2}_{s}+\mathbf{q}^{2}v^{2}_{F})\right)\right]
\left[2+u\left(2\Delta^{2}+\frac{\Omega^{2}_{s}+v^{2}\mathbf{q}^{2}v^{2}_{F}}{2}
\right)\right]\right.\nonumber\\
&\times&\left.\theta_4\left[\frac{(1+v)\Omega_s}{4\pi T}\,,e^{-1/(4uT^2)}\right]
-\theta_4\left[0\,,e^{-1/(4uT^2)}\right]e^{-u\Delta^{2}}-1\right\}.
\label{finiteTB0-polarization}
\ea
The integration over $u$ in (\ref{finiteTB0-polarization}) can be performed explicitly using
a series representation for theta functions, we get in terms of
the integration variable $x=(1+v)/2$:
\ba
\Pi(i\Omega_{s},\mathbf{q})=-\frac{e^2N_f}{2\pi v^{2}_{F}}\int\limits_{0}^1d x
\left[\frac{\Omega^{2}_{s}+\mathbf{q}^{2}v^{2}_{F}+4[\Delta^{2}-x(1-x)
\mathbf{q}^{2}v^{2}_{F}]}{4a(x)}
\frac{\sinh(a(x)/T)}{D(x)}+4T\log\frac{\cosh(\Delta/2T)}{2D(x)}\right],
\ea
where
\ba
a(x)=\sqrt{\Delta^{2}+x(1-x)(\Omega^{2}_{s}+\mathbf{q}^{2}v^{2}_{F})},\quad
D(x)=\cosh^{2}(a(x)/2T)-\sin^{2}(\pi sx).\nonumber
\ea
This expression can be rewritten in somewhat different form if we integrate by parts
the last term in square brackets and then use the  identity among the integrals,
\ba
4T\Omega_{s}\int\limits_{0}^{1}dx\ln[4D(x)]=2\Omega_{s}\int\limits_{0}^{1}dx\frac{a(x)\sinh(a(x)/T)}{D(x)}+
(\Omega^{2}_{s}+\mathbf{q}^{2}v^{2}_{F})\int\limits_{0}^{1}dx
(1-2x)\frac{\sin(2\pi sx)}{D(x)},
\ea
which can be obtained following the method described in the appendix A of
Ref.\cite{Dorey1992NPB}. Finally, we have
\be
\Pi(i\Omega_{s},\mathbf{q})=\frac{e^2N_f}{2\pi}
\frac{\mathbf{q}^{2}}{\Omega^{2}_{s}+\mathbf{q}^{2}v^{2}_{F}}\int\limits_{0}^1d x
\left[2T\log[4D(x)]-\frac{\Delta^{2}}{a(x)}\frac{\sinh(a(x)/T)}{D(x)}\right].
\label{pol-operator-B=0}
\ee
For $\Delta=0$, Eq.(\ref{pol-operator-B=0}) is in agreement with Eq.(A20) (together with (A23), (A26))
in \cite{Dorey1992NPB}
while for $T=0$ it reduces to the well known expression for
the vacuum polarization operator in QED3 \cite{Appelquist}.


\begin{thebibliography}{99}

\bibitem{Novoselov2004S} K.S.~Novoselov, A.K.~Geim, S.V.~Morozov, D.~Jiang,
Y.~Zhang, S.V.~Dubonos, I.V.~Grigorieva and A.A.~Firsov, Science {\bf306},
666 (2004).

\bibitem{Semenoff1984}
G.W.~Semenoff, Phys. Rev. Lett. {\bf53}, 2449 (1984);
D.P.~DiVincenzo and E.J.~Mele, Phys. Rev. B {\bf29}, 1685 (1984).

\bibitem{AndoPRB2002} Y. Zheng and T. Ando, Phys.
Rev. B {\bf65}, 245420 (2002).
\bibitem{GusyninPRL2005} V.P. Gusynin and S.G. Sharapov, Phys.
Rev. Lett. {\bf 95}, 146801 (2005).

\bibitem{PeresPRB2006}N.M.R. Peres, F. Guinea, and A.H. Castro Neto, Phys. Rev. B {\bf
73}, 125411 (2006).

\bibitem{QHE-experiment}K.S.~Novoselov, A.K.~Geim, S.V.~Morozov, D.~Jiang,
M.I.~Katsnelson, I.V.~Grigorieva, S.V.~Dubonos and A.A.~Firsov, Nature {\bf438},
197 (2005);
Y.~Zhang, Y.-W.~Tan, H.L.~Stormer and P.~Kim, Nature {\bf438}, 201 (2005).

\bibitem{universal-optical-theory}V.P.~Gusynin, S.G. Sharapov, and J.~P.~Carbotte,
Phys. Rev. Lett. {\bf96}, 256802 (2006);
L.A.~Falkovsky and A.A.~Varlamov,
Eur. Phys. J. B {\bf56}, 281 (2007).

\bibitem{universal-optical-exp} F.~Wang, Y.~Zhang, C.~Tian, C.~Girit, A.~Zettl,
M.~Crommie, Y. R.~Shen, Science {\bf320}, 206 (2008);
R.R.~Nair, P.~Blake, A.N.~Grigorenko, K.S.~Novoselov, T.J.~Booth, T.~Stauber,
N.M.R.~Peres, A.K.~Geim, Science {\bf320}, 1308 (2008);
Z.Q.~Li, E.A.~Henriksen, Z.~Jiang, Z.~Hao, M.C.~Martin, P.~Kim, H.L.~Storme,
and D.N.~Basov, Nature Physics, {\bf4}, 532 (2008);
K.F.~Mak, M.Y.~Sfeir, Y.~Wu, C.H.~Lui, J.A.~Misewich, and T.F.~Heinz, Phys. Rev. Lett.
{\bf101} 196405 (2008).

\bibitem{magneto-spectroscopy-theory}V.P.~Gusynin and S.G.~Sharapov, Phys. Rev.
B {\bf 73}, 245411 (2006);
V.~P.~Gusynin, S.G.~Sharapov, and J.~P.~Carbotte, Phys. Rev. Lett. {\bf98}, 157402 (2007);
V.~P.~Gusynin, S.G.~Sharapov, and J.~P.~Carbotte, J. Phys.:Cond. Mat. {\bf19}, 026222 (2007).

\bibitem{magneto-spectroscopy-exp}M.L.~Sadowski, G.~Martinez, M.~Potemski, C.~Berger,
and W.A. de Heer, Phys. Rev. Lett. {\bf97},  266405 (2006);
Z.~Jiang, E.A.~Henriksen, L.C.~Tung, Y.-J.~Wang, M.E.~Schwartz, M.Y.~Han, P.~Kim,
and H.L.~Stormer, Phys. Rev. Lett. {\bf98}, 197403 (2007); M.~Orlita and M.~Potemski,
Semicond. Sci. Technol. {\bf25}, 063001 (2010);
I.~Crassee , J.~Levallois , A.~L.~Walter , M.~Ostler , A.~Bostwick,
E.~Rotenberg , T.~Seyller , Dirk van der Marel , A.~B.~Kuzmenko, arxiv:1007.5286.

\bibitem{Gonzalez} J. Gonz$\acute{a}$lez, F. Guinea, and M.A.H. Vozmediano,
 Nucl. Phys. B {\bf 424}, 595 (1994).

\bibitem{Khveshchenko2001PRL} D.V.~Khveshchenko, Phys. Rev. Lett. {\bf87}, 206401 (2001).
\bibitem{Gorbar2002PRB}E.~V.~Gorbar, V.P.~Gusynin, V.~A.~Miransky and I.~A.~Shovkovy,
Phys. Rev. B {\bf66}, 045108 (2002).

\bibitem{instability}O.V.~Gamayun, E.V.~Gorbar, and V.P.~Gusynin, Phys. Rev. B {\bf80},
165429 (2009); {\bf81}, 075429 (2010);
J.E.~Drut and T.A.~L$\ddot{a}$hde,
Phys. Rev. Lett. {\bf102}, 026802 (2009);  Phys. Rev. B {\bf79},
241405(R) (2009); W.~Armour, S.~Hands, and C.~Strouthos, Phys. Rev. B {81}
125105 (2010); J.~Wang, H.A.~Fertig, and G.~Murthy, Phys. Rev. Lett.
{\bf104}, 186401 (2010); J.~Sabio, F.~Sols, and F.~Guinea, Phys. Rev. B {\bf81}, 045428 (2010);
B {\bf82}, 121413 (2010).

\bibitem{footnote}In the presence of a magnetic field, the critical coupling for onset
of a gap generation $g_{c}=0$ due to the magnetic catalysis phenomenon \cite{magcatalysis}.
A non-zero gap leads to divergent resistance at the Dirac point in graphene in
a high magnetic field \cite{Ong2009}.

\bibitem{Wunsch2006NJP} B.~Wunsch, T.~Stauber, F.~Sols and F.~Guinea, New~J.~Phys. {\bf8}, 318 (2006);
E.~H.~Hwang and S.~Das~Sarma, Phys. Rev. B {\bf75}, 205418 (2007).

\bibitem{Pyatkovskiy2009JPCM} P.~K.~Pyatkovskiy, J. Phys.: Condens. Matter {\bf21}, 025506 (2009);
A.~Qaiumzadeh and R.~Asgari, Phys. Rev. B {\bf79}, 075414 (2009).

\bibitem{Shizuya2007PRB} K.~Shizuya, Phys. Rev. B {\bf75}, 245417 (2007).

\bibitem{Roldan} R.~Rold\'an, J.-N.~Fuchs, and M.~O.~Goerbig, Phys. Rev. B {\bf80}, 085408 (2009);
R.~Rold\'an, M.~O.~Goerbig, and J.-N.~Fuchs, Semicond. Sci. Technol. {\bf25}, 034005 (2010).

\bibitem{substrateind-gap}G.~Giovannetti, P.A.~Khomyakov, G.~Brocks, P.J.~Kelly,
and J. van den Brink, Phys. Rev. B {\bf76}, 073103 (2007); S.Y. Zhou, G.-H. Gweon,
A.V. Fedorov, P.N. First, W.A. de Heer, D.-H.~Lee, F.~Guinea, A.H.~Castro Neto,
and A.~Lanzara, Nature Materials {\bf6}, 770 (2007)

\bibitem{Chodos1990PRD} A.~Chodos, K.~Everding, and D.~A.~Owen, Phys. Rev. D {\bf42}, 2881 (1990).

\bibitem{Tahir2008JPCM} M.~Tahir and K.~Sabeeh, J. Phys.: Condens. Matter {\bf20}, 425202 (2008).

\bibitem{Berman2008PRB} O.~L.~Berman, G.~Gumbs and Yu.~E.~Lozovik, Phys. Rev. B {\bf78}, 085401 (2008).

\bibitem{Ando1982RMP} T.~Ando, A~B.~Fowler, and F.~Stern, Rev. Mod. Phys. {\bf54}, 437 (1982).

\bibitem{Davies} J.~H.~Davies, {\it The Physics of Low-dimensional Semiconductors: an Introduction}
(Cambridge University Press, Cambridge, 1998).

\bibitem{Sharapov2004PRB} S.~G.~Sharapov, V.~P.~Gusynin, and H.~Beck, Phys. Rev. B {\bf69}, 075104 (2004).

\bibitem{Shpagin1996} A.~V.~Shpagin, arXiv:hep-ph/9611412 (unpublished).

\bibitem{Gradshteyn} I.~S.~Gradshteyn and I.~M.~Ryzhik,
{\it Tables of Integrals, Series and Products} (Academic Press, New York, 1965).

\bibitem{magcatalysis}V.P.~Gusynin, V.A.~Miransky, and I.A.~Shovkovy,
Phys. Rev. Lett. {\bf 73}, 3499 (1994);

V.~P.~Gusynin, V.~A.~Miransky, S.G.~Sharapov, and I.~A.~Shovkovy,
Phys. Rev. B {\bf74}, 195429 (2006).

\bibitem{Ong2009} J.G.~Checkelsky, L.~Li, and N.P.~Ong,
Phys. Rev. B {\bf 79}, 115434 (2009).

\bibitem{Erdelyi}H.~Bateman and A.~Erdelyi, Higher transcendental functions, V.2,Mc Graw-Hill Co.,
N.Y., 1953.

\bibitem{Dorey1992NPB} N.~Dorey and N.~E.~Mavromatos, Nucl. Phys. B {\bf386}, 614 (1992).

\bibitem{Appelquist}R.~D.~Pisarski, Phys. Rev. D {\bf29}, 2423 (1984);
T.~W.~Appelquist, M.~Bowick, D.~Karabali, and L.~C.~R.~Wijewardhana,
Phys. Rev. D {\bf33}, 3704 (1986).

\end{thebibliography}
\end{document}